\documentclass[doublecol,linenumbers]{epl2} 

\usepackage{graphicx}
\usepackage{dcolumn}
\usepackage{bm}
\usepackage{amsmath}
\usepackage{amssymb}
\usepackage{color}

\renewcommand{\vec}[1]{\mathbf{#1}}

\newif\ifgraph

\graphtrue             %

\title{Diffusion of Chiral Janus Particles in a Sinusoidal Channel}

\author{Xue Ao\inst{1} \and P.~K.~Ghosh\inst{2} \and Y. Li\inst{3}\thanks{E-mail: \email{yunyunli@tongji.edu.cn}} \and G. Schmid\inst{1} \and P. H\"anggi\inst{1,3} \and F.~Marchesoni\inst{3,4}}

\institute{
  \inst{1} Institut f\"ur Physik, Universit\"at Augsburg,
D-86135 Augsburg, Germany\\
\inst{2} Department of Chemistry, Presidency University,
Kolkata - 700073, India\\
\inst{3}Center for Phononics and Thermal Energy Science,
School of Physics Sciences and Engineering, Tongji University,
Shanghai 200092, People's Republic of China\\
\inst{4}Dipartimento di Fisica, Universit\`{a} di
Camerino, I-62032 Camerino, Italy}

\pacs{82.70.Dd}{Colloids}
\pacs{87.15.hj}{Transport dynamics}
\pacs{36.40.Wa}{Charged clusters}

\date{\today}

\abstract{
We investigate the transport diffusivity of artificial microswimmers,
a.k.a. Janus particles, moving in a sinusoidal channel in the absence
of external biases. Their diffusion constant turns out to be quite
sensitive to the self-propulsion mechanism and the geometry of the
channel compartments. Our analysis thus suggests how to best control
the diffusion of active Brownian motion in confined geometries.
}

\begin{document}
\maketitle
\section{Introduction}
Over the last decade the problem of controlling transport of regular
Brownian particles in narrow corrugated channels has attracted the
attention of many investigators with the purpose of better
understanding biological processes in the cell or designing
artificial micro- and nano-devices \cite{ChemPhysChem,RMP2009}. In a
recent development \cite{MSshort} regular Brownian particles have
been replaced with a special type of diffusive tracers, namely, with
active or self-propelled artificial micro-swimmers. Since such
particles operate by harvesting energy from their environment, mostly
in a non-equilibrium steady state, their autonomous transport is
generally enhanced \cite{MSshort}.

Self-propulsion is the ability of most living organisms to move, in
the absence of external drives, thanks to an ``engine'' of their
own~\cite{Purcell}. Optimizing self-propulsion of micro- and
nano-particles (artificial microswimmers) is a growing topic of
today's nanotechnology~\cite{Schweitzer,Rama,Rama2,Ebeling}.
Recently, a new type of artificial microswimmers has been synthesized
\cite{Granick,Chen}, where self-propulsion takes advantage of the
local gradients asymmetric particles can generate in the presence of
an external energy source (self-phoretic effects). Such particles,
called Janus particles (JP), consist of two distinct ``faces", only
one of which is chemically or physically active. Thanks to their
functional asymmetry, JP's can induce either concentration gradients
(self-diffusiophoresis) by catalyzing a chemical reaction on their
active surface~\cite{Paxton1,Bechinger}, or thermal gradients
(self-thermophoresis), e.g., by inhomogeneous light absorption
\cite{Sano} or magnetic excitation~\cite{ASCNano2013JM}.

\begin{figure}[ht]
\centering
\includegraphics[height=8 truecm,angle=0]{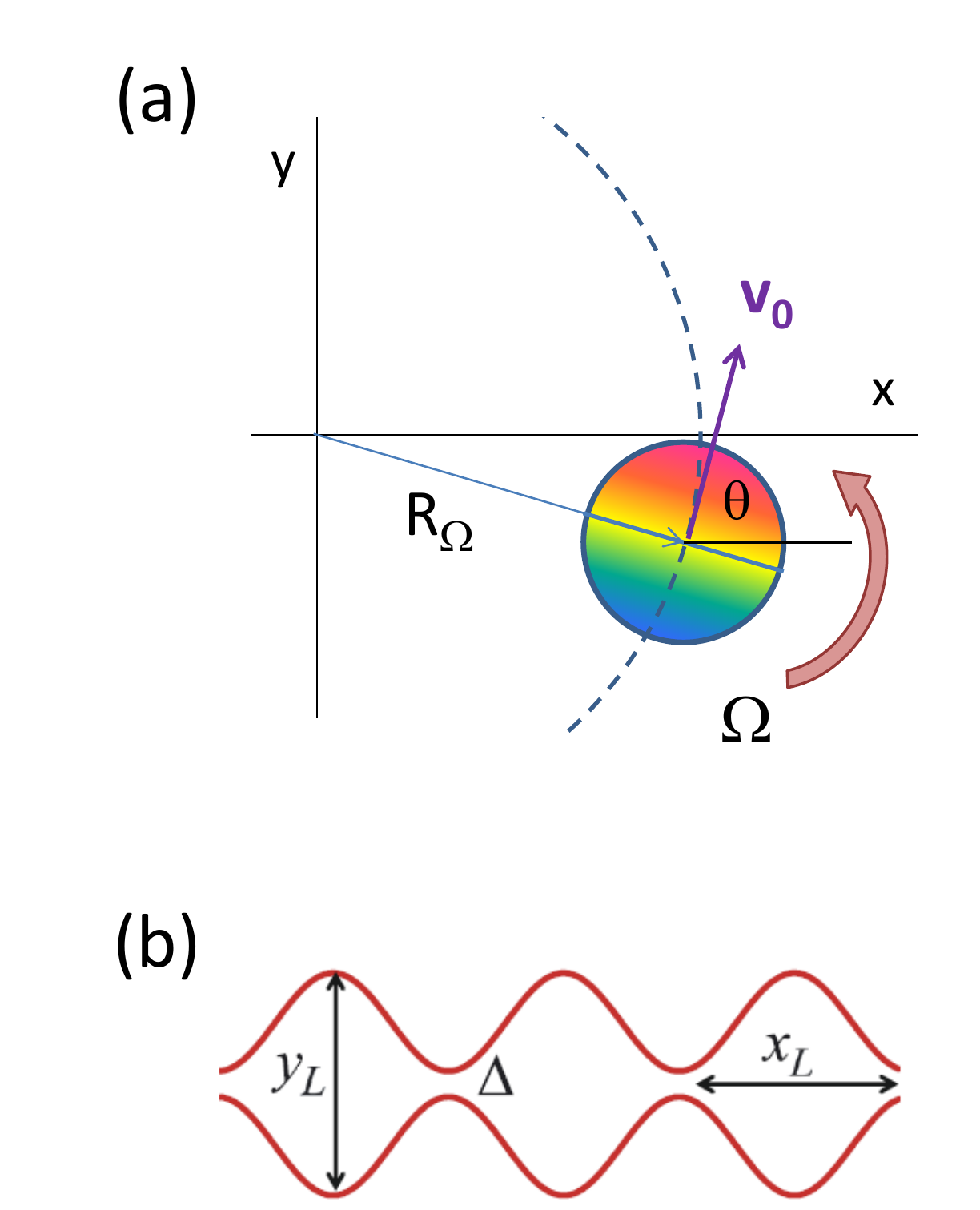}
\caption{(Color online) (a) Chiral levogyre Janus particle with
$\Omega>0$ in the bulk. Sketch of a noiseless particle with
self-propulsion velocity ${\vec v}_0$ and finite torque frequency
$\Omega$, Eq. (\ref{LE}), moving along a circular arc of radius
$R_\Omega$ (dashed line); (b) Sketch of the sinusoidal channel of Eq.
(\ref{wx}). Due to its symmetry, this channel does not rectify JP
diffusion.} \label{F1}
\end{figure}

A self-propulsion mechanism acts on an pointlike particle by means of
a force and, possibly, a torque. In the absence of a torque, the line
of motion is directed parallel to the self-phoretic force and the JP
propels itself along a straight line, until it changes direction
after a mean persistence length, $l_\theta$, due to gradient
fluctuations~\cite{Sen_propulsion} or random collisions against other
particles or geometric boundaries~\cite{Vicsek}. In the presence of
asymmetries in the propulsion mechanism, the self-phoretic force and
the line of motion are no longer aligned and the microswimmer tends
to execute circular orbits with radius $R_\Omega$, as if subject to a
torque with chiral frequency $\Omega$ \cite{Lowen,Julicher} (Fig.
\ref{F1}).

Active chiral motion has long been known in biology
\cite{Brokaw,Julicher,Volpe} and more recently observed in
asymmetrically propelled micro- and nano-rods: A torque can be
intrinsic to the propulsion mechanism, due to the presence of
geometrical asymmetries in the particle fabrication, engineered or
accidental (asymmetric JP's)~\cite{LowenKumm,Ibele,composite}, or
externally applied, for instance, by laser irradiation \cite{Sano} or
hydrodynamic fields~\cite{Stark}.

Active Brownian motion is time correlated {\it per se}, which means
that rectification of a JP can be easily achieved by choosing
spatially asymmetric channel boundaries \cite{MSshort}. As we intend
to investigate the interplay of propulsion chirality and geometric
confinement on the diffusivity of a channeled JP, here we restrict
our analysis to the case of sinusoidal channels, where the
rectification current is known to be identically zero, both for
passive and active Brownian motion. The extension of the present work
to the case of spatially asymmetric channels will be presented in a
forthcoming publication \cite{EPJST}. The main results presented
below can be summarized as follows: (i) A finite torque,
$|\Omega|>0$, tends to suppress the particle diffusivity even in the
bulk, according to a simple phenomenological law that fits remarkably
well the simulation data. This effect grows prominent for chiral
radii much shorter than the self-propulsion length, $R_\Omega \ll
l_\theta$; (ii) The diffusivity of channeled microswimmers, besides
decreasing with $|\Omega|$ as in the bulk, exhibits an additional
side peak, which corresponds to the optimal condition, when a channel
compartment can accommodate for a closed orbit of the chiral swimmer,
thus trapping it; (iii) These properties are rather sensitive to both
the self-propulsion mechanism of the microswimmer and the geometry of
the channel, which points to simple techniques for sorting out
microswimmers according to their swimming properties.

\section{Model}
In order to avoid unessential complications, we restrict
our analysis to the case of 2D channels and pointlike artificial
microswimmers of the JP type \cite{Granick}. The extension of our
conclusions to 3D channels and finite-size particles
\cite{finitesize} is straightforward. A chiral JP gets a continuous
push from the suspension fluid, which in the overdamped regime
amounts to a rotating self-propulsion velocity ${\vec v_0}$ with
constant modulus $v_0$ and angular velocity $\Omega$. Additionally,
the self-propulsion direction varies randomly with time constant
$\tau_\theta$, under the combined action of thermal noise and
orientational fluctuations intrinsic to the self-propulsion
mechanism.

\begin{figure}[tp]
\centering
\includegraphics[width=0.46\textwidth]{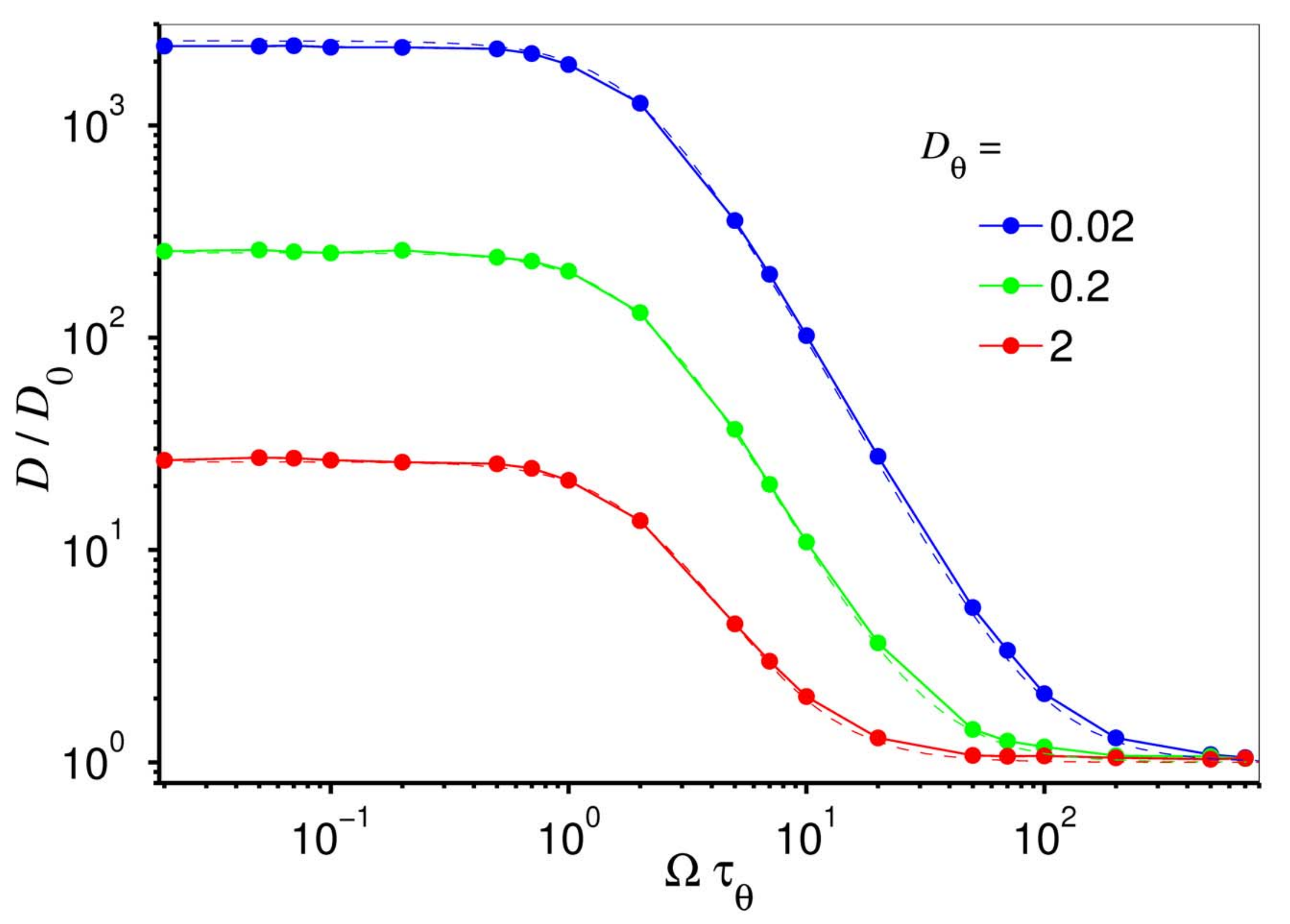}
\caption{(Color online) Diffusion of a levogyre JP with $\Omega \geq
0$ and $v_0=1$ in a straight channel: $D_{\rm ch}$ vs. $\Omega$ for
different $D_\theta$ and $D_0=0.01$.  The boundaries $w_\pm(x)$ are
given by Eq. (\ref{wx}) with $\Delta=y_L=1$. The dashed curves
represent the corresponding phenomenological law, $D_{\rm ch}=D$,
holding for straight channels, with $D$ given in Eq. (\ref{diffW}).
Our results are independent on the sign of $\Omega$ and the width of
the straight channel (not shown).} \label{F2}
\end{figure}

The bulk dynamics of such an overdamped chiral JP obeys the Langevin
equations \cite{Lowen,Volpe,SoftMatter}
\begin{eqnarray}
\label{LE} \dot x &=& v_0\cos \theta +\xi_x(t) \\ \nonumber \dot y
&=& v_0\sin \theta +\xi_y(t) \\ \nonumber \dot \theta &=&\Omega
+\xi_\theta(t),
\end{eqnarray}
where the coordinates of the particle center of mass, ${\bf
r}=(x,y)$, are subject to the Gaussian noises $\xi_{i}(t)$, with
$\langle \xi_{i}(t)\rangle=0$ and $\langle
\xi_{i}(t)\xi_{j}(0)\rangle=2D_0\delta_{ij}\delta (t)$ for $i=x,y$,
modeling the equilibrium thermal fluctuations in the suspension
fluid. The channel is directed along the $x$ axis, the
self-propulsion velocity is oriented at an angle $\theta$ with
respect to it and the sign of $\Omega$ is chosen so as to coincide
respectively with the positive (levogyre) and negative (dextrogyre)
chirality of the swimmer, see Fig. \ref{F1}. The orientational
fluctuations of the propulsion velocity are modeled by the Gaussian
noise $\xi_\theta(t)$ with $\langle \xi_{\theta}(t)\rangle=0$ and
$\langle \xi_{\theta}(t)\xi_{\theta}(0)\rangle=2D_{\theta}\delta(t)$,
where, as shown below, $D_\theta$ sets the orientational time
constant, $\tau_\theta$, of the self-propulsion velocity,
$\tau_{\theta}=2/D_{\theta}$. Accordingly, the microswimmer mean free
self-propulsion path approximates a circular arc of radius
$R_\Omega=v_0/|\Omega|$ and length $l_\theta=v_0\tau_\theta$
\cite{Lowen}. Therefore, for $R_\Omega \lesssim l_\theta$, or
equivalently, $|\Omega|\tau_{\theta} \gtrsim 1$ (strong chirality
regime), chiral effects tend to appreciably suppress the ensuing
active Brownian diffusion as shown below.

All noise sources in Eq. (\ref{LE}) have been treated as
independently tunable, although, strictly speaking, thermal and
orientational fluctuations may be statistically correlated depending
on the self-propulsion mechanism \cite{Schweitzer,Paxton1,Bechinger}.
Moreover, we ignored hydrodynamic effects, which are known to favor
clustering in dense mixtures of JP's
\cite{Ripoll,Buttinoni,Marchetti} and even cause their capture by the
channel walls \cite{Takagi}. However, both effects are negligible for
low density mixtures of pointlike spherical JP's. Moreover, we made
sure that the parameters used in our simulations were experimentally
accessible, as apparent on expressing times in seconds and lengths in
microns and comparing with the experimental setups of Refs.
\cite{Bechinger,Volpe}.

When confined to a channel directed along the $x$ axis, the particle
transverse coordinate, $y$, is bounded between a lower and upper
wall, $w_{-}(x)\leq y \leq w_{+}(x)$, with
\begin{eqnarray}
\label{wx} w_\pm(x) &=& \pm \frac{1}{2} \left [\Delta
+(y_L-\Delta)\sin^2\left(\frac{\pi}{x_L}x\right) \right ],
\end{eqnarray}
Such a sinusoidal channel is periodic; its compartments have length
$x_L$ and are mirror symmetric under both coordinate inversions, $x
\to -x$ and $y \to -y$, i.e., centro-symmetric. Throughout our
analysis we assumed that the width, $\Delta$, of the pores connecting
the compartments are much narrower than the maximum channel
cross-section, i.e., $\Delta \ll y_L$.

Simulating a constrained JP requires defining its collisional
dynamics at the boundaries. For the translational velocity $\vec{\dot
r}$ we assumed elastic reflection. Regarding the coordinate $\theta$,
we assumed that it does not change upon collision (sliding b.c.
\cite{MSshort}). As a consequence the active particle slides along
the walls for an average time of the order of $\tau_\theta$, until
the $\theta$ fluctuations, $\xi_\theta(t)$, redirect it toward the
interior of the compartment. In the limit of strong persistency of
the propulsion mechanism, $l_\theta \gg x_L, y_L$, and weak
chirality, $R_\Omega \gg l_\theta$ the stationary particle
probability density $P(x,y)$ accumulates along the boundaries; this
effect is the strongest in the noiseless case, $D_0=0$
\cite{MSshort}.

The dispersion of a Brownian particle along the channel
axis~\cite{Machura} is an important issue experimentalists address
when trying to demonstrate rectification. Indeed, drift currents, no
matter how weak, can be detected over an affordable observation time
only if the relevant dispersion is sufficiently small. This issue is
of paramount importance when one handles with active Brownian
particles, like JP', whose stochastic dynamics is characterized by
strong persistency, or long correlation times. Under such conditions
the current literature on classical diffusion is of little help
\cite{ChemPhysChem,Brenner}.  To this purpose we computed the
transport diffusivity, $D_{\rm ch}$, of a JP in the sinusoidal
channel of Eq. (\ref{wx}), as the limit
\begin{equation} \label{diffch}
D_{\rm ch}=\lim_{t\to \infty}[\langle x^2(t)\rangle - \langle x(t)
\rangle^2]/(2t),
\end{equation}
which we checked to exist for all simulation parameters (normal
diffusion limit).

\begin{figure}
\centering
\includegraphics[width=0.46\textwidth]{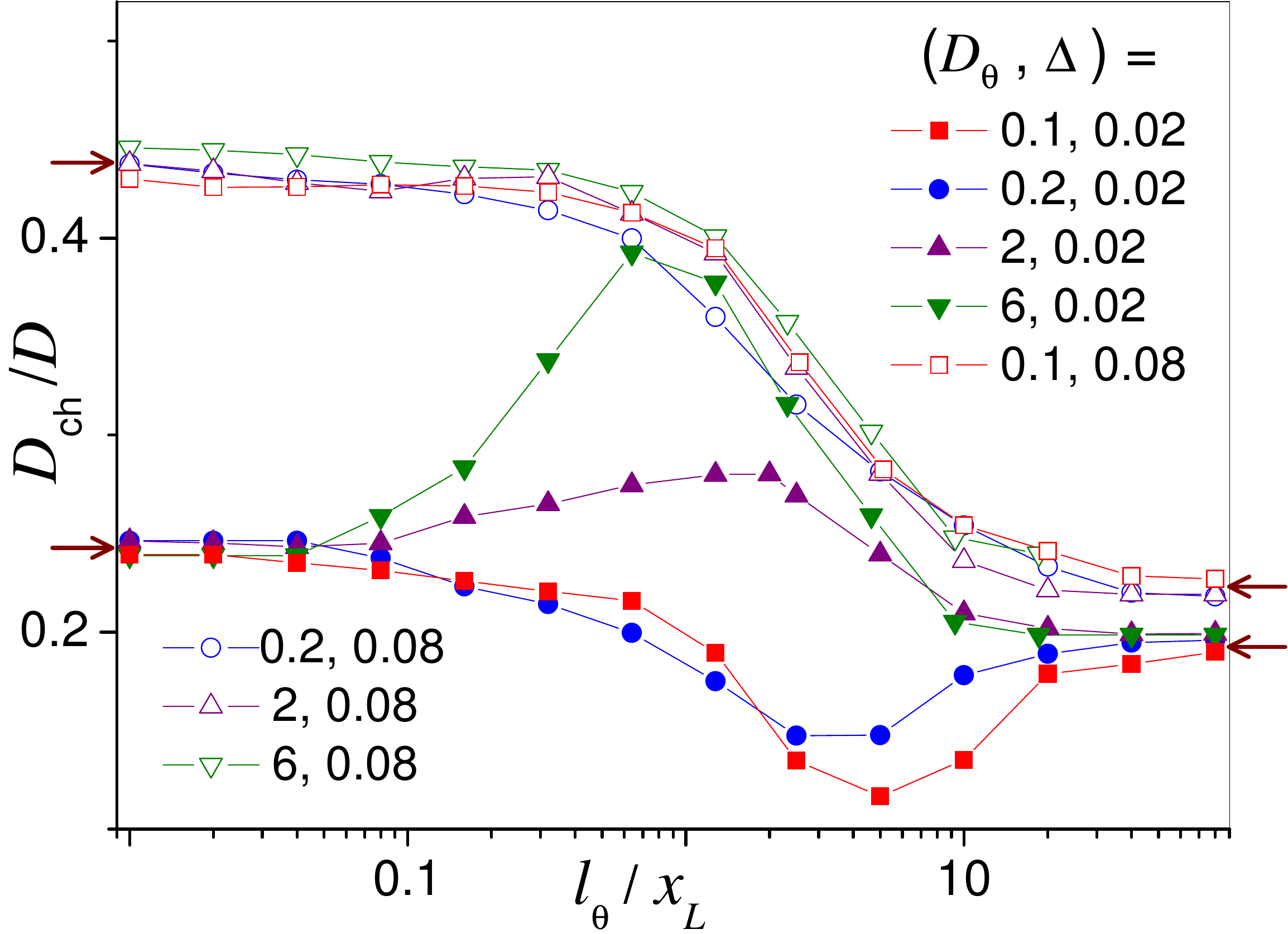}
\caption{(Color online) Diffusion of a nonchiral JP in the sinusoidal
channel of Eq. (\ref{wx}): $D_{\rm ch}/D$ vs. $v_0$ at constant
$D_\theta$ and $\Delta$ (see legend). $D$ is the bulk diffusivity of
Eq. (\ref{diff1}). Other simulation parameters are $x_L=y_L=1$ and
$D_0=0.05$. The left and right arrows denote, respectively, the
estimated values of the suppression constants, $\kappa_0$ and
$\kappa_s$, introduced in the text, i.e., $\kappa_0=0.25$ and
$\kappa_s=0.20$ for $\Delta=0.02$; $\kappa_0=0.45$ and
$\kappa_s=0.23$ for $\Delta=0.08$.} \label{F3}
\end{figure}

\section{Bulk diffusion, $D$}
An analytical solution of the model Eq.
(\ref{LE}) is out of question even in the bulk (i.e., in the absence
of boundaries) and for $\Omega=0$ ({\it nonchiral} JP). However, on
noticing that \cite{MSshort,Marchetti}
\begin{equation} \label{eta}
\langle \cos \theta (t) \cos \theta (0) \rangle=\langle \sin \theta
(t) \sin \theta (0) \rangle=(1/2)e^{-|t|D_\theta},
\end{equation}
and the first two LE's of Eq. (\ref{LE}) are statistically
independent, namely $\lim_{t\to \infty}\langle \cos \theta (t) \sin
\theta (t) \rangle=0$, one concludes immediately that a nonchiral
particle diffuses according to F\"urth's law
\begin{eqnarray}
\label{furth} \langle \Delta x(t)^2\rangle &=& \langle \Delta
y(t)^2\rangle \\ \nonumber &=& 2 (D_0+v_0^2\tau_\theta/4)t
+(v_0^2\tau_\theta^2/2)(e^{-2t/\tau_\theta}-1).
\end{eqnarray}
For $t \gg \tau_\theta$ we thus recover the asymptotic normal
diffusion law, $\langle \Delta x(t)^2\rangle = 2Dt$, where the
constant $D$ apparently consists of two distinct contributions,
\begin{equation}
\label{diff1} D=D_0+D_s,
\end{equation}
due to the randomness of, respectively, the thermal fluctuations,
$D_0$, and self-propulsion, $D_{s}= {v_0^2\tau_\theta}/{4}$.

Determining the $\Omega$ dependence of the bulk diffusivity,
$D(\Omega)$, of a chiral JP is a more challenging task. Our
prediction is the phenomenological law
\begin{equation}
\label{diffW} D(\Omega)=D_0+\frac{D_s}{1+ (\Omega \tau_\theta/2)^2},
\end{equation}
where $D(0)$ coincides with $D$ in Eq. (\ref{diffch}). The derivation
of this law can be summarized as follows: (i) As suggested by the
separation between thermal and active diffusion in Eqs. (\ref{diff1})
and (\ref{diffW}), we focused on the limiting case of zero thermal
noise, $D_0=0$; (ii) In view of the identities in Eq. (\ref{eta}), we
noticed that the ensuing velocity components $\dot x$ and $\dot y$
can be regarded as two independent non-Gaussian exponentially
correlated noises with zero mean, intensity $D_s$ and correlation
time $\tau_\theta$; (iii) Finally, having set $\xi_x(t)=\xi_y(t)=0$,
we took the time derivative of the first two LE's in Eq. (\ref{LE})
and linearized them as
\begin{eqnarray}
\ddot x&=&-\Omega \dot y- 2\dot x/\tau_\theta +2\eta(t)/\tau_\theta \label{deriva}\\
\nonumber \ddot y&=&+\Omega \dot x- 2\dot y/\tau_\theta
+2\eta(t)/\tau_\theta,
\end{eqnarray}
where $\eta(t)$ denotes a stationary white Gaussian noise with zero
mean and strength $D_s$; (iv) From the approximate 2D stationary
dynamics of Eq. (\ref{deriva}) our prediction for $D(\Omega)$ in Eq.
(\ref{diffW}) follows suite \cite{Taylor,Kur} (for more details see
Ref. \cite{thesis}).

Owing to the b.c. adopted here, for a JP diffusing in a straight
channel, say, with $w_\pm(x)=\pm y_L/2$, bulk and channel diffusivity
coincide, $D_{\rm ch}=D$. This statement is confirmed by the fact
that the simulation curves displayed in Fig. \ref{F2} do not depend
on $y_L$. Most remarkably, all three curves are closely fitted by the
phenomenological law (\ref{diffW}). As expected from Eq.
(\ref{diffW}), $D(\Omega)$ interpolates the diffusivity of a
nonchiral JP, Eq. (\ref{diff1}), at $\Omega=0$ and the thermal
diffusivity, $D_0$, for $\Omega \to \infty$. In the latter limit,
i.e., for $R_\Omega/l_\theta \to 0$, self-diffusion is totally
suppressed.

\begin{figure}
\centering
\includegraphics[width=0.46\textwidth]{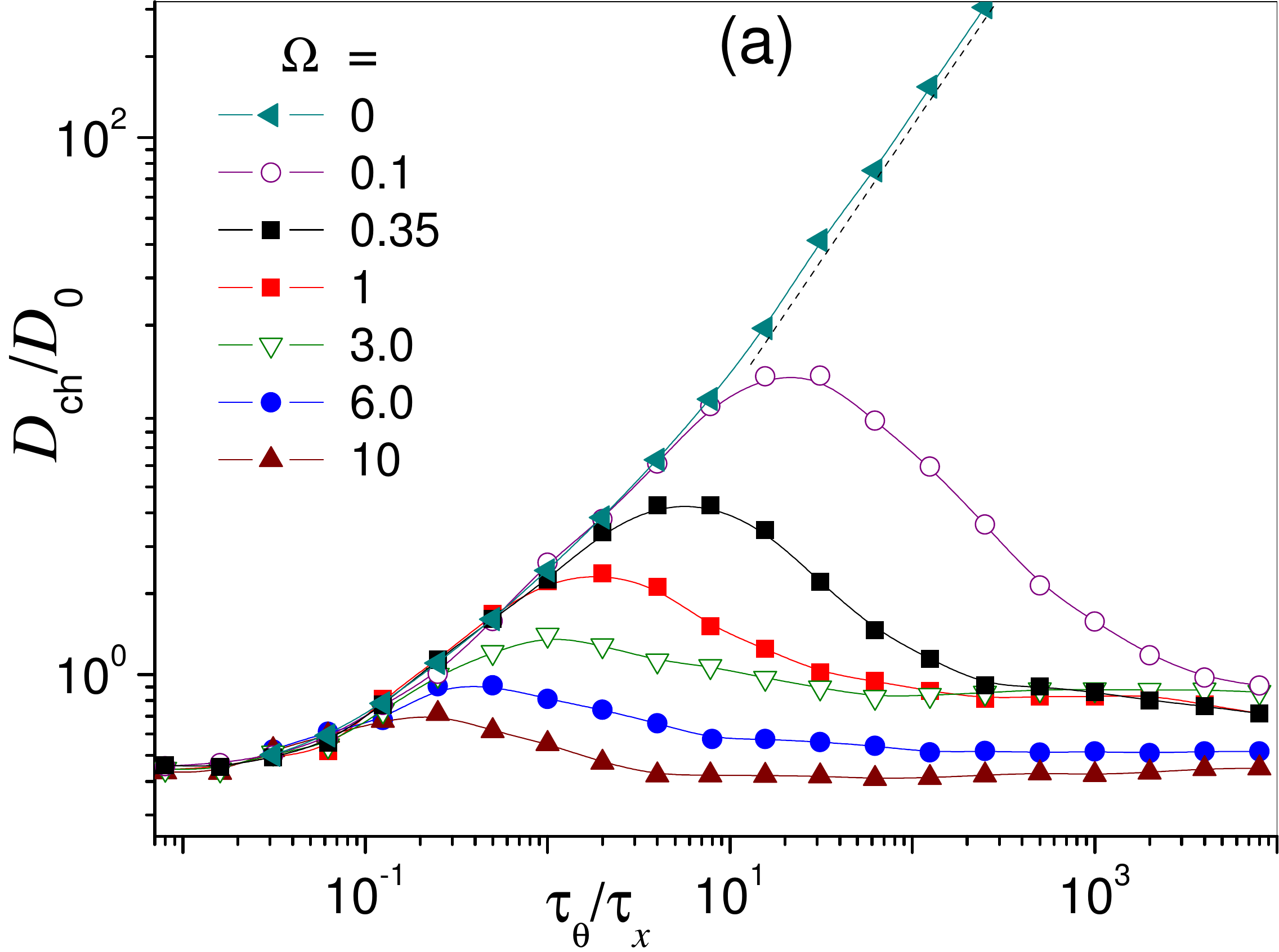}
\includegraphics[width=0.46\textwidth]{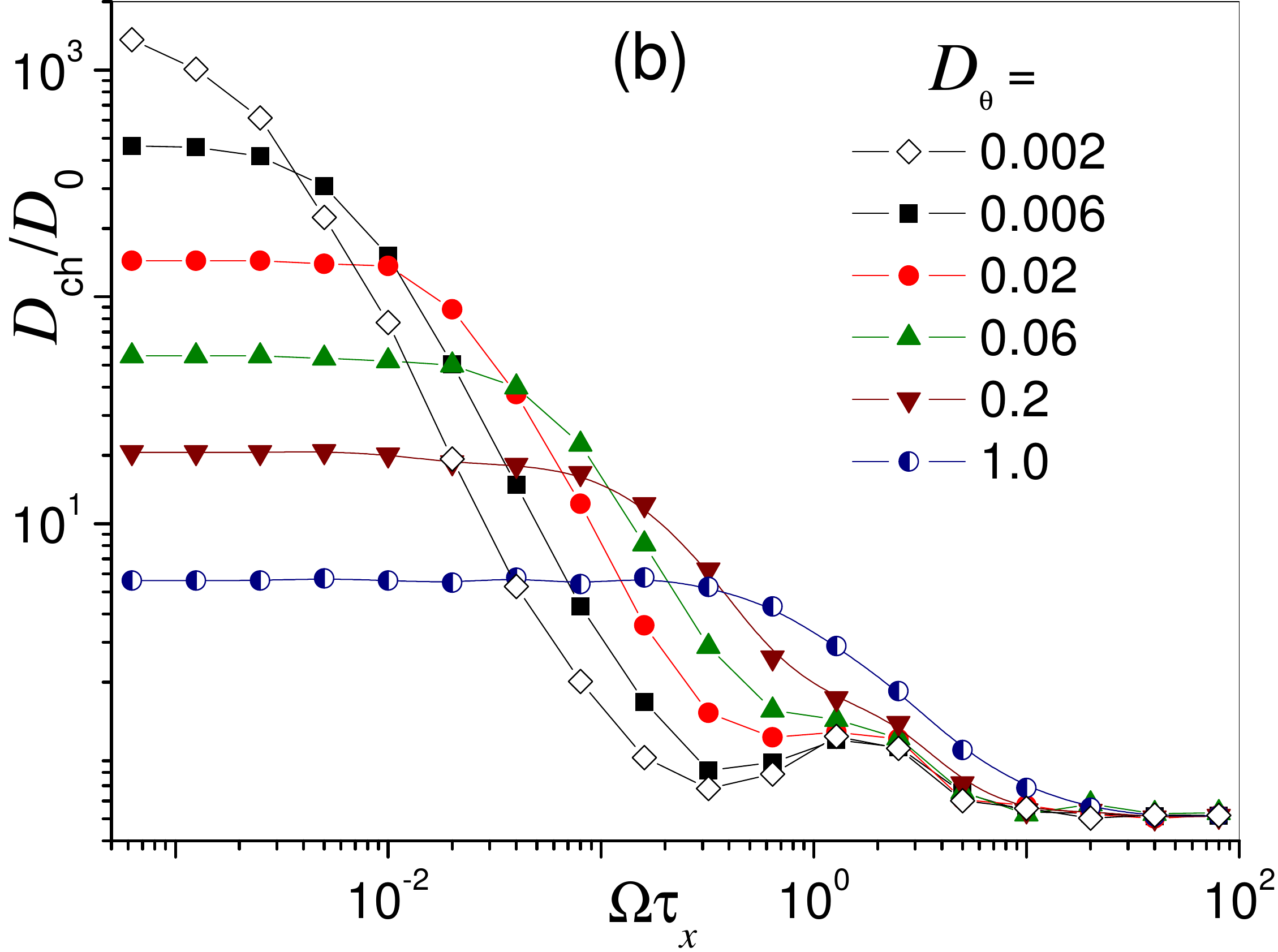}
\caption{(Color online) (a) Diffusion of a levogyre JP in the
sinusoidal channel of Eq. (\ref{wx}): (a) $D_{\rm ch}/D_0$ vs.
$\tau_\theta$ for different $\Omega$. Notice that the smaller
$\Omega$, the slower is convergence of $D_{\rm ch}/D_0$ to $\kappa_0$
for large $\tau_theta$; (b) $D_{\rm ch}/D_0$ vs. $\Omega$ for
different $D_\theta$. Here, $\tau_x\equiv x_L/v_0$, $v_0=1$,
$D_0=0.05$, and $\Delta=0.08$. The dashed line in (a) represents the
asymptotic linear power law of Eq.(\ref{diff1}). } \label{F4}
\end{figure}

\section{Channel diffusion, $D_{\rm ch}$}
When confined to a corrugated
channel, the particle diffusivity is suppressed by the geometric
constrictions represented by the pores, as shown in Figs. \ref{F3}
and \ref{F4}, respectively, for nonchiral and chiral JP's diffusing
along a sinusoidal channel.

In the absence of self-propulsion, say, for $v_0=0$ (or,
equivalently, $l_\theta=0$), the bulk diffusivity is $D=D_0$, see Eq.
(\ref{diff1}), and the channel diffusivity can be written as $D_{\rm
ch}=\kappa_0 D_0$, with $\kappa_0$ a well studied function of
$\Delta$ and $D_0$ \cite{Schmid,Bosi}. In the opposite limit of
strong self-propulsion, $v_0 \to \infty$, the bulk diffusion of a
{\it nonchiral} JP is governed by self-diffusion, that is, $D\simeq
D_s$ and, accordingly, in the channel $D_{\rm ch}=\kappa_s D_s$.
Both limits of $D_{\rm ch}$ are illustrated in Fig. \ref{F3} for
different values of $\Delta$ and $D_\theta$. 
Apparently, neither $\kappa_0$ nor $\kappa_s$ depend on $D_\theta$
and both are smaller than one. This conclusion applies to different
compartment geometries, symmetric and asymmetric, alike, as confirmed
by further simulation results reported in Ref. \cite{EPJST}.

The mechanisms underlying the suppression of channel diffusion
quantified by the constants $\kappa_0$ and $\kappa_s$, are different.
For a regular Brownian particle with $v_0=0$ moving in a narrow
channel, $\kappa_0$ can be estimated in Fick-Jacobs' approximation
for smoothly corrugated channels \cite{ChemPhysChem,Schmid} and in
mean-first-exit time formalism for sharply compartmentalized channels
\cite{Bosi,Borromeo1}. In both cases, $\kappa_0$ strongly depends on
the compartment volume, the pore width and thermal noise, since
particle diffusion mostly happens away from the walls. For nonchiral
self-propelling microswimmers with $l_\theta \gg x_L,y_L$, the
constant $\kappa_s$ is mostly determined by the b.c. introduced to
model the particle collisions against the channel walls. For sliding
b.c., the probability flows (consequence of the JP's piling up
against the boundaries \cite{SoftMatter}) are modulated by the wall
profiles, $w_\pm(x)$, and thus not much sensitive to the pore width
itself (as long as the particle size is negligible; see Ref.
\cite{EPJST} for more details). The distinct $\Delta$ dependence of
$\kappa_0$ and $\kappa_s$ is apparent in Fig. \ref{F3}.

We consider next the case of channeled {\it chiral} JP's. The
diffusivity of a levogyre JP's in a sinusoidal channel, illustrated
in Fig. \ref{F4}, clearly points to two different chirality-induced
suppression mechanisms. We have already shown how chirality limits
the bulk diffusion of JP's with long self-propulsion time constants,
that is $|\Omega|\tau_\theta \gg 1$ or $R_\Omega \ll l_\theta$, see
Eq. (\ref{diffW}). On the other hand, when the chiral radius
$R_\Omega$ grows smaller than the compartment dimensions, say,
$R_\Omega \ll x_L$, or $|\Omega|\tau_x \gg 1$, all swimmers, even
those with long self-propulsion length, $l_\theta \gg x_L$, are
expected to perform closed orbits and thus get trapped inside the
channel compartments \cite{SoftMatter}. Such a geometric condition is
likely to produce an additional suppression of channel diffusion.

With these premises the dependence of $D_{\rm ch}$ on the
self-propulsion mechanism parameters, $\tau_\theta$ and $\Omega$, can
be satisfactorily explained, at least, at a qualitative level. Curves
of $D_{\rm ch}$ versus $\tau_\theta$ at constant $\Omega$ are
reported in Fig. \ref{F4}(a). By inspection one notices immediately
that: (i) At large $\tau_\theta$, $D_0$ is negligible with respect to
$D_s$ and, again, $D_{\rm ch} \simeq \kappa_s D(\Omega)$, like for
nonchiral JP's. Most remarkably, we checked that $\kappa_s$ does not
sensibly depend on $\Omega$; (ii) The curves $D_{\rm ch}$ versus
$\tau_\theta$ go through a maximum, $D_{\rm ch}^{\rm max} \simeq
\kappa_s D_s/2$, the position of which, $|\Omega|\tau_\theta=2$, is
insensitive to the channel geometry. Indeed, position and height of
the maxima located at $\tau_\theta > \tau_x$ can be closely
approximated by plotting the bulk diffusivity of Eq. (\ref{diff1}) as
a function of $\tau_\theta$ and making use of the relation $D_{\rm
ch}=\kappa_s D$; (iii) For finite $\Omega$, active diffusivity in the
channel is suppressed both for $\tau_\theta \to 0$ and $\tau_\theta
\to \infty$. In both limits, one thus expects that $D_{\rm ch}
(\Omega) \to \kappa_0 D_0$, where $\kappa_0$ is the corresponding
suppression constant at $\Omega=0$ given in Fig. \ref{F3}. Note that
at low $\Omega$ the convergence toward the expected large
$\tau_\theta$ asymptote is very slow. The same asymptote is
approached by the large $\Omega$ tails of the curves plotted in Fig.
\ref{F4}(b).

Moreover, the $\Omega$ dependence of $D_{\rm ch}$ at constant
$\tau_\theta$, illustrated in Fig. \ref{F4}(b), shows explicitly that
(iv) $D_{\rm ch}$ starts decreasing appreciably with $\Omega$ only
for $|\Omega|\tau_\theta \gtrsim 2$, that is in coincidence with the
maxima displayed in  Fig. \ref{F4}(a); (v) Small diffusivity peaks
emerge for $|\Omega|\tau_\theta \gg 1$. They are centered around a
certain value of $\Omega$, $\Omega_M$, which does not depend on the
time constant $\tau_\theta$. $\Omega_M$ can be estimated by noticing
that on increasing $\Omega$ the chiral radius $R_\Omega=v_0/|\Omega|$
decreases, until the microswimmer performs a full circular orbit
inside the channel compartment, without touching its walls (actually
a logarithmic spiral with exponentially small steps \cite{Lowen}). In
the noiseless limit, this happens for $2R_\Omega \simeq x_L$, that
is, $\Omega_M \simeq 2v_0/x_L$. This condition can be regarded as the
onset of a mechanism of dynamical trapping. In Brownian transport
theory, the onset of a trapping mechanism generally corresponds to an
excess diffusion peak \cite{peak0,peak1,peak2}: That is precisely the
phenomenon we see at work here. Of course, this argument requires
that $\Omega_M \tau_\theta \gg 1$, to ensure a sufficiently long
self-propulsion time. Both our estimate for $\Omega_M$ and the
condition for the diffusivity peak to appear are in close agreement
with the data displayed in Fig. \ref{F4}(b).

\section{Conclusions}
We numerically investigated the diffusion of
artificial active microswimmers moving along narrow periodically
corrugated channels. Our work is meant to complement the earlier
literature on the rectification of active microswimmers in confined
geometries. Transport quantifiers, like rectification power and
diffusivity, strongly depend on the particle self-propulsion
mechanism and the channel compartment geometry. The emerging picture
suggests the possibility of developing new control techniques for the
manipulation of artificial microswimmers, which are well within the
reach of today's technology. Specialized microfluidic circuits can be
designed, for instance, to guide chiral microswimmers to a designated
target. The same technique can be utilized to fabricate monodisperse
chiral microswimmers (presently a challenging technological task). By
the same token, microswimmers capable of inverting chirality upon
binding to a load, can operate as chiral shuttles along a suitably
corrugated channel even in the absence of gradients of any kind.

\acknowledgments
X.A. has been supported by the grant Equal Opportunity for Women in
Research and Teaching of the Augsburg University. P.H. and G.S.
acknowledge support from the cluster of excellence Nanosystems
Initiative Munich (NIM). Y.L. was supported by the NSF China under
grants No. 11347216 and 11334007, and by Tongji University under
grant No. 2013KJ025. F.M. thanks the Alexander von Humboldt Stiftung
for a Research Award.


\begin{thebibliography}{0}

\bibitem{ChemPhysChem}
\Name{Burada P.~S., H\"anggi P., Marchesoni F., Schmid G. \and
Talkner P.} for a review see,
  \REVIEW{ChemPhysChem} {10} {2009} {45}.

\bibitem{RMP2009}
\Name{H\"{a}nggi P. \and Marchesoni F.}
  \REVIEW{Rev. Mod. Phys.} {81}{2009} {387}.

\bibitem{MSshort}
\Name{ Ghosh P.~K., Misko V.~R., Marchesoni F. \and Nori F.}
  \REVIEW{Phys. Rev. Lett.}{110}{2013}{268301}.
%
\bibitem{Purcell}
\Name{Purcell E.~M.}
  \REVIEW{Am. J. Phys.}{45}{1977}{3}.
%
\bibitem{Schweitzer}
\Name{Schweitzer F.}
\Book{Brownian Agents and Active Particles}
  \Publ{Springer, Berlin Heidelberg}
  \Year{2003}.
%
\bibitem{Rama}
\Name{Ramaswamy S.}
  \REVIEW{Annu. Rev. Condens. Matter Phys.}{1}{2010}{323}.
  %
\bibitem{Rama2}
\Name{Vicsek T.  \and Zafeiris A.}
  \REVIEW{Phys. Rep.} {517}{2012}{71}.
%
\bibitem{Ebeling}
\Name{Romanczuk P., B\"ar M., Ebeling W., Lindner B.  \and Schimansky-Geier L.}
  \REVIEW{Eur. Phys. J. Special Topics}{202}{2012}{1}.
%
\bibitem{Granick}
  \Editor{Jiang S. \and Granick S.}
  \Book{Janus Particle Synthesis, Self-Assembly and Applications}
    \Publ{RSC Publishing, Cambridge}
    \Year{2012}.
%
\bibitem{Chen}
\Name{Walther A. \and M\"uller A.~H.~E.}
  \REVIEW{Chem. Rev.}{113}{2013}{ 5194}.
%
\bibitem{Paxton1}
\Name{Paxton W.~F., Sundararajan S., Mallouk T.~E. \and Sen A.}
  \REVIEW{Angew. Chem. Int. Ed.}{45}{2006}{5420}.
%
%
\bibitem{Bechinger}
\Name{Volpe G., Buttinoni I., Vogt D., K\"{u}mmerer H.-J. \and Bechinger C.}
  \REVIEW{Soft Matter}{7}{2011}{8810}.
%
\bibitem{Sano}
\Name{Jiang H.~R., Yoshinaga N. \and Sano M.}
  \REVIEW{Phys. Rev. Lett.}{105}{2010}{268302}.
%
\bibitem{ASCNano2013JM}
\Name{Baraban L., Streubel R., Makarov D., Han L., Karnaushenko D., Schmidt  O.~G. \and Cuniberti G.}
  \REVIEW{ACS Nano}{7}{2013}{1360}.
%
\bibitem{Sen_propulsion}
\Name{Hong Y., Velegol D., Chaturvedi N. \and  Sen A.}
  \REVIEW{Phys. Chem. Chem. Phys.} {12}{2010} {1823}.
%
\bibitem{Vicsek}
\Name{ B\'uz\'as A.,  Kelemen L.,  Mathesz A., Oroszi L., Vizsnyiczai G. , Vicsek T. \and Ormos P.}
  \REVIEW{Appl. Phys. Lett.} {101}{2012}{041111}.

\bibitem{Lowen}
\Name{van Teeffelen S. \and L\"owen H.}
  \REVIEW{Phys. Rev. E} {78}{2008}{020101}.

\bibitem{Julicher}
\Name{Friedrich B.~M. \and J\"ulicher F.}
  \REVIEW{Phys. Rev. Lett.} {103}{2009}{068102}.

\bibitem{Brokaw}
\Name{Brokaw C.~J.}
  \REVIEW{J. Exp. Biol.} {35}{1958}{97}.
  \REVIEW{J. Cell. Comp. Physiol.}{54}{1959}{95}.

\bibitem{Volpe}
\Name{Mijalkov M. \and Volpe G.}
  \REVIEW{Soft Matter} {9}{2013} {6376}.

\bibitem{LowenKumm}
\Name{K\"ummel F., ten Hagen B., Wittkowski R., Buttinoni I., Eichhorn R., Volpe G., L\"owen  H. \and Bechinger C.}
  \REVIEW{Phys. Rev. Lett.} {110}{2013}{198302}.

\bibitem{composite}
\Name{Boymelgreen A., Yossifon G., Park S. \and Miloh T.}
  \REVIEW{Phys. Rev. E}{89}{2014}{011003(R)}.

\bibitem{Ibele}
\Name{Sen A., Ibele M., Hong Y. \and Velegol D.}
  \REVIEW{Faraday Discuss.} {143}{2009}{15}.

\bibitem{Stark}
\Name{Z\"ottl A. \and Stark H.}
  \REVIEW{Phys. Rev. Lett.}{108}{2012}{218104}.
%
%
%

\bibitem{EPJST}
\Name{Ao X., Ghosh P.~K., Li Y., Schmid G., H\"anggi P. \and Marchesoni F.}
  \REVIEW{Eur. Phys. J Special Topics}{223}{2014}{3227}.
%
\bibitem{finitesize}
\Name{Ten Hagen B., van Teeffelen S. \and L\"{o}wen H.}
  \REVIEW{J. Phys.: Condens. Matter} {23}{2011}{194119}.
%
\bibitem{SoftMatter}
\Name{Li Y., Ghosh P.~K., Marchesoni F. \and Li B.}
  \REVIEW{Phys. Rev. E}{90}{2014}{062301}.
%
\bibitem{Ripoll}
\Name{Ripoll M., Holmqvist P., Winkler R.~G., Gompper G., Dhont J.~K.~G. \and Lettinga M.~P.}
  \REVIEW{Phys. Rev. Lett.} {101}{2008}{168302}.
%
\bibitem{Buttinoni}
\Name{Buttinoni I., Bialk\`e J., K\"ummel F.,  L\"owen H, Bechinger C. \and Speck T.}
  \REVIEW{Phys. Rev. Lett.} {110}{2013}{ 238301}.
%
\bibitem{Marchetti}
\Name{Fily Y. \and Marchetti M.~C.}
  \REVIEW{Phys. Rev. Lett.}{108}{2012}{235702}.
%
\bibitem{Takagi}
\Name{Takagi D., Palacci J., Braunschweig A.~B., Shelley M.~J. \and Zhang J.}
  \REVIEW{Soft Matter} {10}{2014}{1784}.
%
\bibitem{Machura}
\Name{Machura L., Kostur M., Talkner P., Luczka J., Marchesoni F. \and H\"anggi P.}
  \REVIEW{Phys. Rev. E} {70}{2004}{061105}.
%
\bibitem{Brenner}
\Name{Brenner H. \and Edwards D.~A.}
\Book{Macrotransport Processes}
\Publ{Butterworth-Heinemann, New York}
\Year{1993}.
%
\bibitem{Taylor}
\Name{Taylor J.~B.}
  \REVIEW{Phys. Rev. Lett.}{6}{1961}{262}.
%
\bibitem{Kur}
\Name{Kur\ifmmode \mbox{\c{s}}\else \c{s}\fi{}uno\ifmmode \check{g}\else
\v{g}\fi{}lu B.}
  \REVIEW{Phys. Rev.} {132}{1963}{21}.
%
\bibitem{thesis}
Ao X., PhD thesis (Augsburg University, in preparation).
%
\bibitem{Schmid}
\Name{Burada P.~S., Schmid G., Reguera D., Rubi J.~M. \and H\"anggi P.}
  \REVIEW{Phys. Rev. E} {75}{2007}{051111}.
%
\bibitem{Bosi}
\Name{Bosi L., Ghosh P.~K. \and Marchesoni F.}
  \REVIEW{J. Chem. Phys.}{137}{2012}{174110}.
%
\bibitem{Borromeo1}
\Name{Borromeo M. \and Marchesoni F.}
  \REVIEW{Chem. Phys.} {375}{2010}{536}.
%
\bibitem{peak0}
\Name{Schreier M., H\"anggi P. \and Pollak E.}
  \REVIEW{Europhys. Lett.} {44}{1998}{416}.
%
\bibitem{peak1}
\Name{Costantini G. \and Marchesoni F.}
  \REVIEW{Europhys. Lett.} {48}{1999}{491}.
%
\bibitem{peak2}
\Name{Reimann P., Van den Broek C., Linke H., H\"anggi P., Rub\'i J. M. \and Perez Madrid A.}
  \REVIEW{Phys. Rev. Lett.} {87}{2001}{010602}.

%
\end{thebibliography}
\end{document}